\documentclass[aip, amsmath,amssymb,reprint]{revtex4-2}
\usepackage[utf8]{inputenc}
\usepackage[T1]{fontenc}
\usepackage{graphicx}
\usepackage{dcolumn, lipsum,graphicx,float}
\usepackage{diagbox}
\graphicspath{ {./Figures/} }
\usepackage{slashbox}
\usepackage{gensymb}
\usepackage{wrapfig}
\usepackage[dvipsnames]{xcolor}
\usepackage{hyperref}
\hypersetup{
    colorlinks=true,
    linkcolor=blue,
    citecolor = blue,
    urlcolor=blue
}

\begin{document}

\title{Simplified Josephson-junction fabrication process for reproducibly high-performance superconducting qubits}

\author{A. Osman}
\email{amr.osman@chalmers.se}
\author{ J. Simon, A. Bengtsson, S. Kosen, P. Krantz, D. Perez, M. Scigliuzzo}
\author{Jonas Bylander}
\author{A. Fadavi Roudsari}

\affiliation{Department of Microtechnology and Nanoscience, Chalmers University of Technology, 412 96 Gothenburg, Sweden}

\date{\today}


\begin{abstract}
We introduce a simplified fabrication technique for Josephson junctions and demonstrate superconducting Xmon qubits with $T_1$ 
relaxation times averaging above $50~\mu$s ($Q>1.5\times 10^6$).
Current shadow-evaporation techniques for aluminum-based Josephson junctions require a separate lithography step to deposit a patch that makes a galvanic, superconducting connection between the junction electrodes and the circuit wiring layer. The patch connection eliminates parasitic junctions, which otherwise contribute significantly to dielectric loss.
In our patch-integrated cross-type (PICT) junction technique, we use one lithography step and one vacuum cycle to evaporate both the junction electrodes and the patch.
In a study of more than 3600 junctions, we show an average resistance variation of 3.7$\%$ on a wafer that contains forty $0.5\times 0.5$-cm$^2$ chips, with junction areas ranging between 0.01 and 0.16~$\mu$m$^2$. The average on-chip spread in resistance is 2.7$\%$, with 20 chips varying between 1.4 and 2$\%$. For the junction sizes used for transmon qubits, we deduce a wafer-level transition-frequency variation of 1.7---2.5$\%$. We show that 60---70$\%$ of this variation is attributed to junction-area fluctuations, while the rest is caused by tunnel-junction inhomogeneity. Such high frequency predictability is a requirement for scaling-up the number of qubits in a quantum computer. 
\end{abstract}

\maketitle

\section*{Introduction}

Superconducting quantum circuits constitute a promising architecture for the realization of quantum computers. Over the past two decades, many researchers have put strong effort into improving the fabrication processes of superconducting circuits to increase the achievable quantum-coherence time. \cite{martinis_decoherence_2005,nersisyan_manufacturing_2019,richardson_2020} 
On the other hand, the reproducibility of Josephson-junction (JJ) fabrication has only recently gained considerable interest, motivated by the need for a scalable process to engineer multiqubit systems. \cite{lokhtov_rep,Bumble_rep,  Krantz2010InvestigationOT,pop_fabrication_2012,MIT_reproducibility_2015, Osman2019,kreikebaum2019improving,hertzberg_laser} 
Variation of the Josephson inductance represents the dominant cause of qubit-frequency variation, e.g.\@ for the transmon-type qubit. \cite{koch_charge-insensitive_2007} An increased reproducibility of JJs is therefore important to ensure predictability of qubit frequencies, in order to enable pulsed-microwave control while avoiding cross-talk, a necessity for scaling up beyond a few coupled qubits. \cite{neill_blueprint_2018, arute_quantum_2019, krantz_quantum_2019} Reproducibility is also important for other superconducting devices, particularly the traveling-wave parametric amplifier (TWPA), \cite{TWPA_mohebbe, TWPA_brien, white_traveling_2015, TWPA_Macklin307, TWPA_zorin, TWPA_zorin_2} which requires impedance matching and identical inductances along the long, lumped-element transmission line to avoid reflections and signal loss.

The all-dominant materials combination of JJs for qubit applications consists of an aluminum/aluminum oxide/aluminum (Al/AlO$_x$/Al) sandwich fabricated by double-angle shadow evaporation of aluminum, within one vacuum cycle, with controlled in-situ oxidation in-between to form the tunneling barrier. \cite{dolan1997, CMOS} 
Maintaining a galvanic, superconducting contact between the JJ's electrodes and the rest of the circuit is important, in order to avoid forming ``parasitic'' junctions in series, whose dielectric loss tangent contributes to decoherence and parameter fluctuations.
\cite{martinis_decoherence_2005,Muller_2019} In fact, Lisenfeld et al. \cite{TLS_KIT} found, in a recent study, that 40$\%$ of the two-level-system (TLS) defects responsible for dielectric loss were located within the parasitic junction formed due to the shadow evaporation technique (with the remaining 60$\%$ located at circuit interfaces and almost none within the JJ itself). Additionally, Nersisyan et al. \cite{nersisyan_manufacturing_2019} showed that the area of this parasitic junction adversely affects the coherence of the qubit. To mitigate this issue, the parasitic junction can be eliminated by depositing a patch (or bandage) layer that connects the junction electrodes to the rest of the circuit, after removal of the native oxide of aluminum. \cite{dunsworth_characterization_2017} 
However, the further processing introduced by adding the patch can introduce more losses, especially those caused by interfacial resist residues that are difficult to remove. \cite{quintana_characterization_2014} 

In this work, we propose and demonstrate a new technique to fabricate both the junction and the patch layer in a single lithography step by evaporating from three angles. We name the technique patch-integrated cross-type (PICT), with reference to the cross-type Josephson junctions first proposed in ref.~\onlinecite{CMOS}. We favorably evaluate the quality and reliability of our process by characterizing both the qubit coherence and the fabrication reproducibility. We measured the $T_1$ relaxation and $T^*_2$ Ramsey free-induction decay times, and their fluctuations, showing an average quality factor of $1.6 \times 10^6$, i.e.\@ without additional losses in comparison to our standard fabrication process. \cite{burnett_decoherence_2019} In addition, we characterized the reproducibility of the JJ parameters by fabricating a statistically
significant number (> 3600) of Josephson junctions
and measuring their normal resistance, $R_N$, at room temperature. $R_N$ is directly proportional to the Josephson inductance, $L_J$, and therefore, the measurement of $R_N$ provides information on reproducibility of the qubit frequency, $f_{01}$. \cite{ambegaokar_tunneling_1963, koch_charge-insensitive_2007, Osman2019} 
The measured inter-chip variation is 3.7$\%$ across a wafer, which drops to an average on-chip value of 2.7$\%$. Furthermore, the resistance spread increases with decreasing junction size: for sizes used in fixed-frequency transmon qubits (0.02---0.06~$\mu$m$^2$), we found a wafer-level variation of 3.4---4.9$\%$, corresponding to 1.7---2.5$\%$ in qubit frequency.

\section*{Method}

\begin{figure}[t]
\centering
    \includegraphics[width=0.96\linewidth]{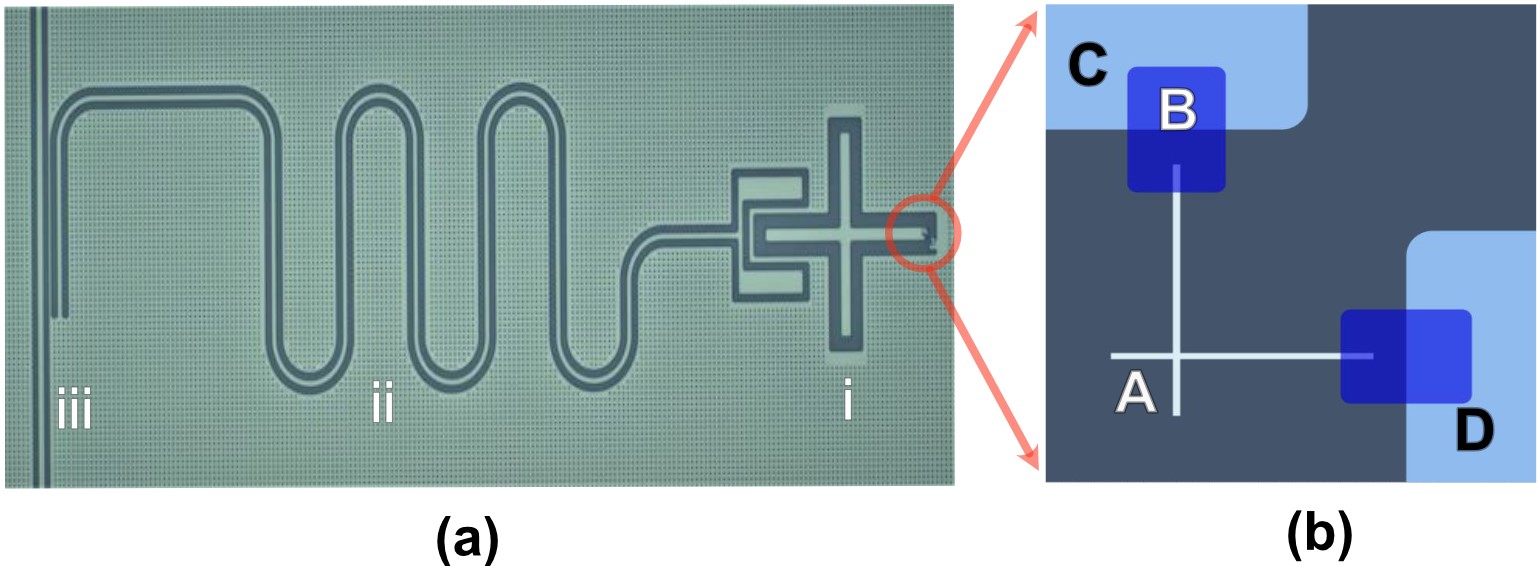}
    \caption{\small{(a) Micrograph of a device, consisting of an Xmon qubit (i) capacitively coupled to a resonator (ii), which is inductively coupled to a transmission line (iii). (b) Layout of the Josephson junction (A) and the patch for the standard process (B), connecting the junction electrodes to the Xmon capacitor (C) and ground (D).}}
    \label{fig:method}
\end{figure}

The process described in this work builds on the background of our previous standard qubit design and fabrication techniques. \cite{burnett_decoherence_2019}
The layout of a typical device is shown in Fig.~\ref{fig:method}(a): it consists of a transmon/Xmon-type qubit \cite{koch_charge-insensitive_2007,barends_coherent_2013} (i) that is capacitively coupled to a readout resonator (ii), which is inductively coupled to a transmission line (iii). In our standard fabrication process, \cite{burnett_decoherence_2019, barends_coherent_2013} the aluminum ground plane is first deposited using electron beam evaporation. The wiring (transmission line, resonator, and shunt capacitor) and the flux trapping holes are then patterned using optical lithography and etched using wet chemistry.
Figure~\ref{fig:method}(b) shows the JJ layout and the bandages or patches commonly used to connect it to the rest of the circuit. \cite{dunsworth_characterization_2017} 
The junction itself (A) is patterned using electron beam lithography (EBL), followed by the cross technique to deposit the Al layers \cite{CMOS} (two thin-film depositions at an angle separated by a $90 \degree$ planetary turn, and oxidation to form the tunneling barrier), in a Plassys MEB 550s evaporator, followed by lift-off. Next, the patch layer (B) is patterned in a final lithography step, which ensures galvanic connection of the junction to the capacitor (C) and the ground plane (D). After development, the oxide layer on top of the aluminum is milled in-situ before the deposition of the patch and lift-off. 

\begin{figure}[t]
\centering
    \includegraphics[width=0.86\linewidth]{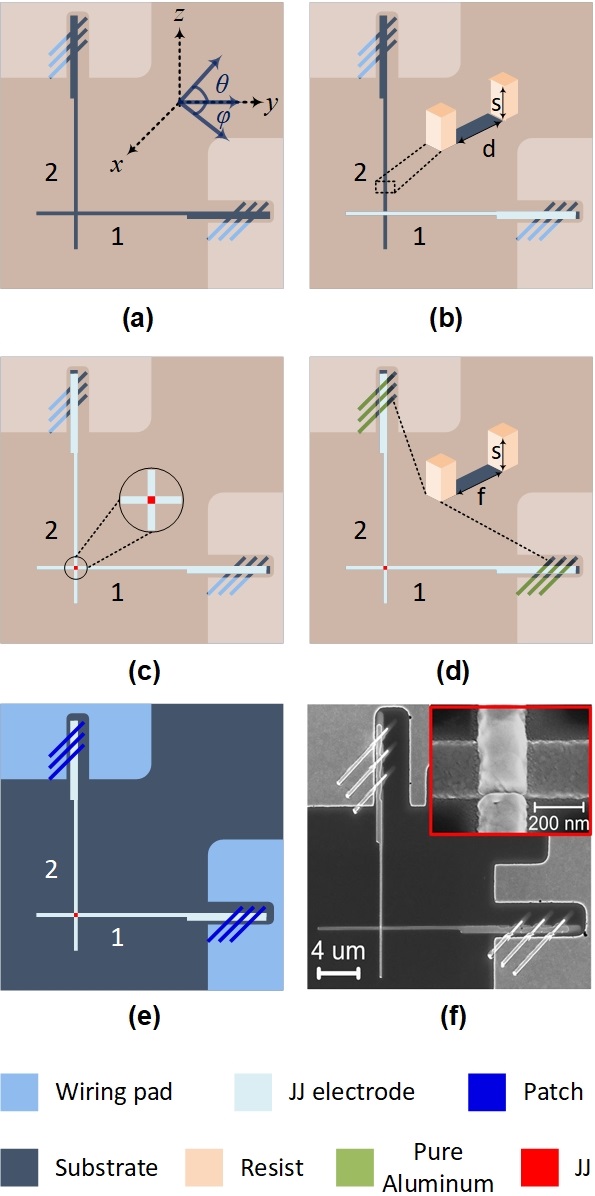}
    \caption{\small{(a)-(e) Schematic of the PICT process flow. $\theta$ and $\varphi$ are the planetary and tilt angles of the sample holder, respectively. Deposition of Al on top of the resist is not shown for clarity. (f) SEM image of the fabricated junction with the patch layer. }}
    \label{fig:process}
\end{figure}

In our PICT process, we pattern both the junction and the patch in one EBL step and evaporate the thin films within one vacuum cycle. A modification to the patch layout makes this possible, as shown in Fig.~\ref{fig:process}, where instead of rectangles, the patches are shaped like $45 \degree$ fringes to provide selective deposition and milling when the resist is thick enough. This eliminates an entire lithography run and reduces the total steps of Josephson-junction fabrication by 50$\%$.

The subsequent evaporation steps are shown in Fig.~\ref{fig:process}(a-e), where $\theta$ and $\varphi$ are the planar and tilt angles (from the y-axis) of the sample holder, respectively. The evaporation and ion milling are both perpendicular to the yz-plane, pointing towards the -x direction. The reference position is shown in Fig.~\ref{fig:process}(a), where $\varphi = \theta = 0\degree$. When the angles are set at $\theta = \theta_1$ and $\varphi = \varphi_1$, 
first, the sample holder turns counter clockwise around the x-axis by $\theta_1$ degrees in the yz-plane. Next, the sample holder turns (tilts) around the z-axis by $\varphi_1$ degrees.

The first junction electrode (1) is deposited at $\theta = 0\degree$ and $\varphi = 45\degree$ (Fig.~\ref{fig:process}(b)) and oxidized to form the tunneling barrier. The second electrode (2) is then deposited at $\theta = -90\degree$ and $\varphi = 45\degree$ in Fig.~\ref{fig:process}(c) and oxidized to form a protective layer for the Al film; this controlled oxidation is preferred over natural oxidation of aluminum as a result of exposure to the ambient. The purpose of the two slits in both the ground and capacitor electrodes, into which the two junction electrodes fit, is to avoid any discontinuity in the deposited electrodes due to shadowing. Having slits is not a general necessity; we added them to keep the electrodes and the wiring layout as close as possible to the design of our standard devices. Next the surface is prepared for patching, i.e.\@ removing the oxide atop the Al films in the fringes.
This is achieved by Ar$^+$ ion milling of the substrate at $\theta = -45\degree$ and $\varphi = 45\degree$ (Fig.~\ref{fig:process}(d)). At this angle, the resist wall protects the junction area from being milled away. Al is then deposited from the same angle in order to form the patch (Fig.~\ref{fig:process}(e)). The Al is anew oxidized to create a protective oxide. Figure~\ref{fig:process}(f) shows a scanning electron micrograph (SEM) of the junction and the patch after lift-off.

Apart from $\varphi$ and $\theta$, three other parameters have to be taken into account in this process:\cite{Osman2019} the resist thickness $s$, the width of a fringe $f$, and the width of the junction electrodes $d$.  The cross-type technique requires that $s > d \tan{\varphi}$  to obtain selective electrode deposition \cite{CMOS} (3D schematic, Fig.~\ref{fig:process}(b)). However, here the more stringent condition $s > d \sqrt{2} \tan{\varphi}$ applies to avoid deposition or milling of the junction area when forming the patch layer. Additionally, it is required that $f < s/(\sqrt{2} \tan{\varphi})$ to avoid Al deposition on the fringes during deposition of the electrodes, assuming a fringe angle of $45 \degree$ (3D schematic on Fig. \ref{fig:process}(d)). For all of these inequalities, $\varphi$ is left variable. In our implementation, $s=0.95~\mu$m, $f=0.4~\mu$m, and $\varphi=45 \degree$.

We note that other patch patterns exist, which can connect the junction electrodes to the rest of the circuit---the key is to shape them such that they are shadowed and protected from the evaporating metal when the junction is being made.

\section*{Coherence characterization}

\begin{table}[b]
\caption{\small{\label{tab:Q_frequency} Parameters of the measured qubits. For each qubit, the junction area is based on the design parameters, while 
the qubit frequency, $f_{01}$, the relaxation time, $T_1$, and the free-induction decay time, $T^*_2$, were measured. Note that $T_1$ and $T^*_2$ are shown as means, given in $\mu$s, plus/minus one standard deviation, given in percent, representing temporal fluctuations around the mean. \cite{burnett_decoherence_2019} }}

\begin{ruledtabular}

\centering
\footnotesize
\begin{tabular}{lccccc}
Qubit & \shortstack{Junction \\ area \\ ($\mu$m$^2$)} &  \shortstack{$f_{01}$ \\ (GHz) } & \shortstack{$T_1$ \\($\mu$s)} & \shortstack{$T^*_2$ \\ ($\mu$s)} & \shortstack{$Q$ \\ ($10^6$)}\\\hline
S-X1 & 0.024 & 4.219 & 54 $\pm$ 23$\%$  &  73 $\pm$ 22$\%$ & 1.4\\
P-X1 & 0.024  & 4.19 & 59 $\pm$ 12$\%$ & 65 $\pm$ 26$\%$ & 1.6\\
S-X2 & 0.0225 & 4.268 & 55 $\pm$ 23$\%$ & 53 $\pm$ 36$\%$ & 1.5\\
P-X2 & 0.0225 & 4.15 & 56 $\pm$ 23$\%$ & 57 $\pm$ 29$\%$ & 1.5\\
S-X3 & 0.021 & 3.956 & 48 $\pm$ 35$\%$ & 41 $\pm$ 37$\%$ & 1.2\\
P-X3 & 0.021 & 3.933  & 69 $\pm$ 20$\%$ & 43 $\pm$ 28$\%$ & 1.7\\
\end{tabular}
\end{ruledtabular}

\end{table}

\begin{figure}[t]
\centering
    \includegraphics[width=0.98\linewidth]{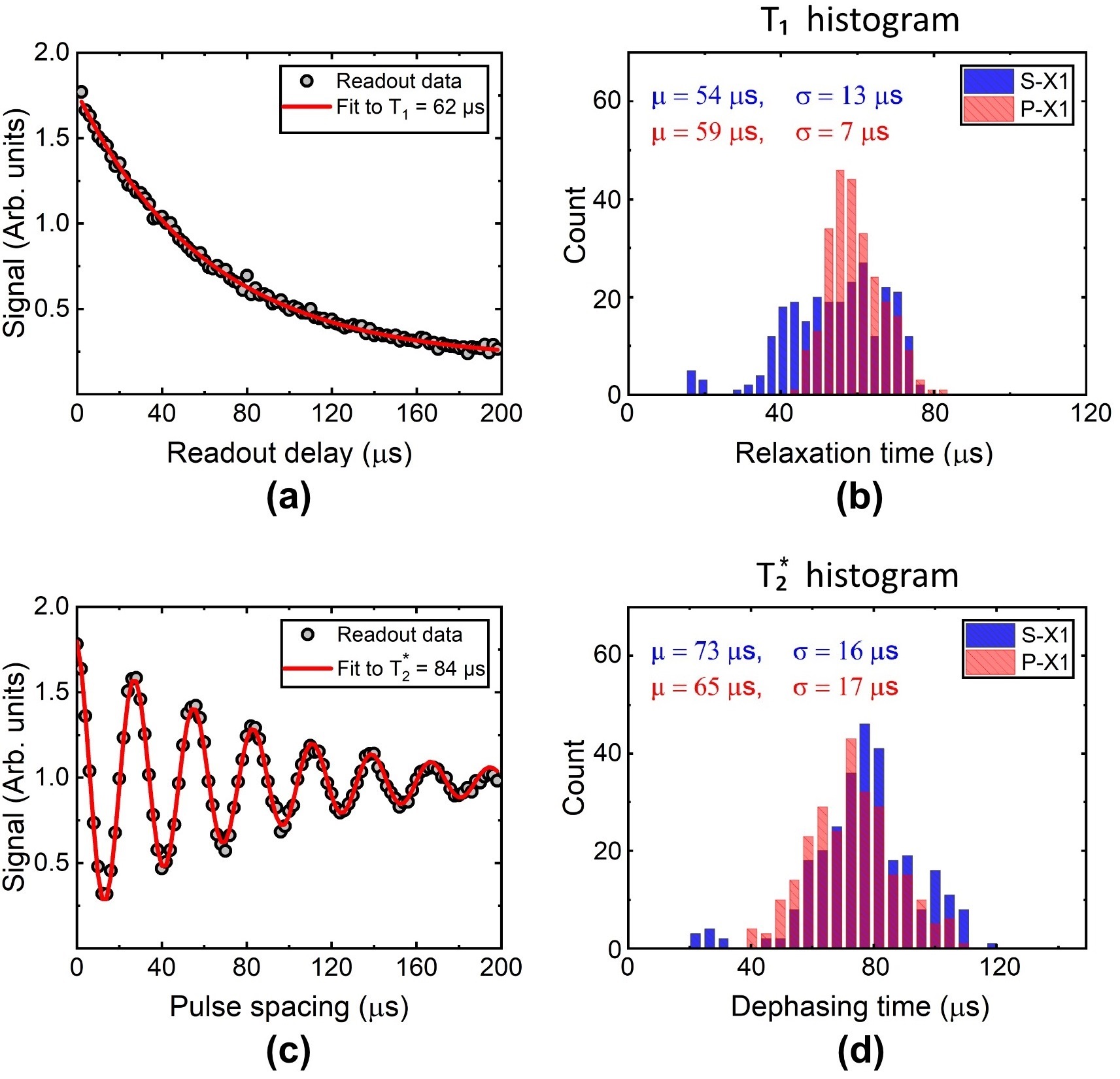}
    \caption{\small{Coherence characterization for two different qubits, S-X1 (standard process) and P-X1 (PICT process). (a) and (c) represent $T_1$ and $T^*_2$ data and fits for one measurement iteration on P-X1. (b) and (d) are histograms of $T_1$ and $T^*_2$ over 250 measurements on both S-X1 and P-X1.}}
    \label{fig:T1_T2}
\end{figure}

In order to quantitatively investigate the quality of the junctions made by the new process, we fabricated Xmon qubits and compared their performance against our benchmark. \cite{burnett_decoherence_2019} Our study involves two chips designated as S (standard) and P (PICT), each containing three Xmons denoted X1, X2, and X3. To establish a fair comparison between the two processes, we fabricated both chips on the same wafer, so that they would undergo the exact same steps for the ground plane and wiring layer until the wafer was split in two for the fabrication of the junctions and the patches. Each chip was packaged and wire bonded in a copper box, mounted onto the mixing chamber of a dilution refrigerator, and measured at a temperature below 12 mK.

Table \ref{tab:Q_frequency} presents the qubit parameters. We find that the frequency of each qubit on the P chip matches that of its pair on the S chip within a few tens of MHz, which indicates that the new process did not cause any large variations in the frequency. We measured $T_1$ and $T_2^*$ for each of the six qubits more than 250 times over a time span of approximately 15 hours in order to capture the statistics of the ubiquitous parameter fluctuations.\cite{burnett_decoherence_2019} Figures~\ref{fig:T1_T2}(a) and (c) show examples of measurements (data points) and their fits (continuous line) on qubit P-X1. Figures~\ref{fig:T1_T2}(b) and (d) show histograms of $T_1$ and $T_2^*$ for qubits S-X1 and P-X1, showing very similar values between the chips. For all the qubits, the values of $T_1$ and $T_2^*$ and their standard deviations are summarized in Table~\ref{tab:Q_frequency}. Since these qubits have different frequencies, we can most fairly compare their performance by rescaling their $T_1$ to the quality factor, $Q = 2\pi f_{01} T_1$. The average $Q$ for the PICT-JJ qubits is 1.6 $\times$ 10$^6$, while for the standard qubits, we obtain a negligibly different number: 1.4 $\times$ 10$^6$.

\begin{figure*}
\centering
    \includegraphics[width=0.94\linewidth]{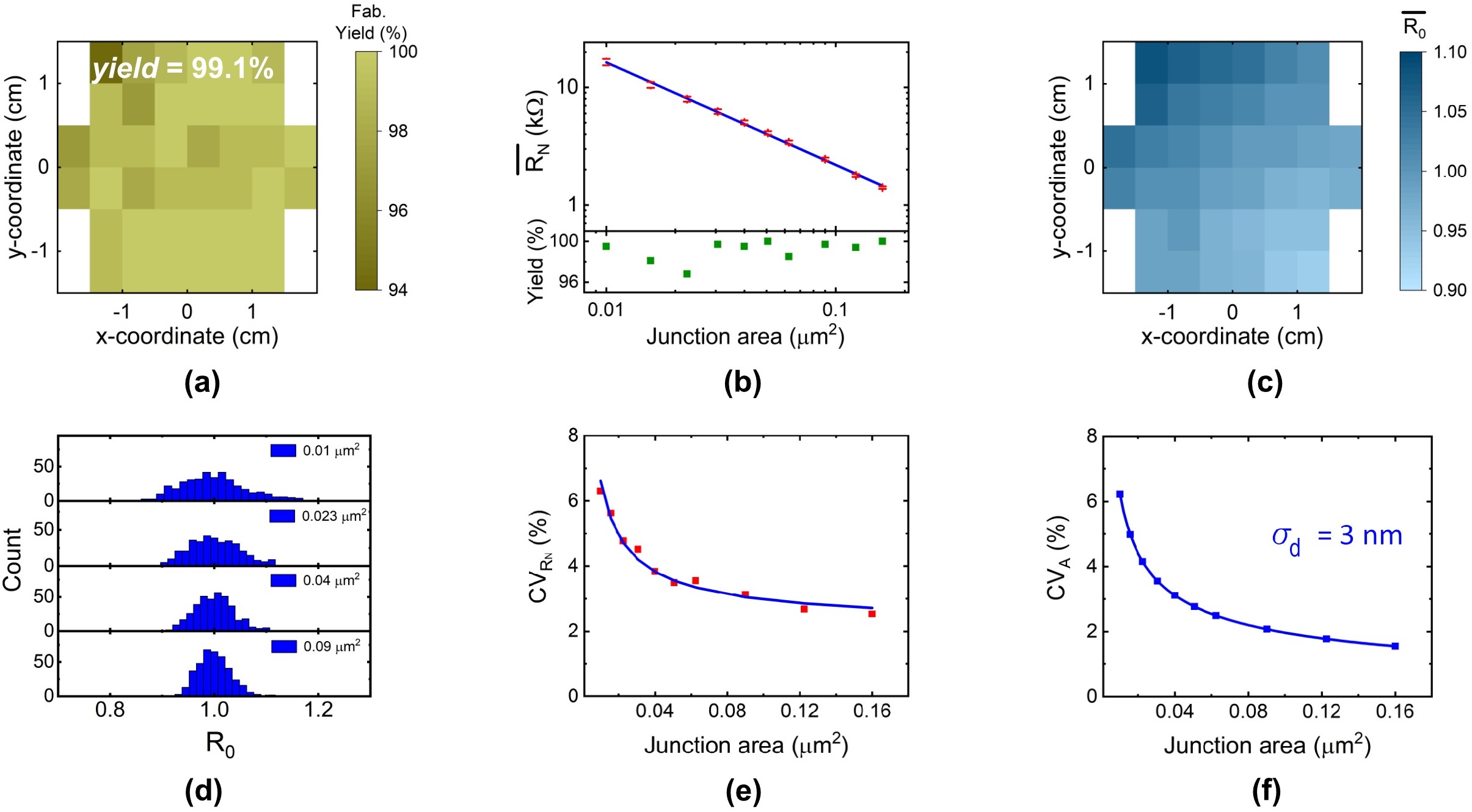}
    \caption{\small{Reproducibility of the PICT junction resistance. (a) Heat map of the fabrication yield for forty $5\times5~\mathrm{mm}^2$ chips across a 76-mm wafer. (b, top) Mean value of $R_N$, with error bars (red), vs.\@ junction area, and a linear fit (blue). (b, bottom) Fabrication yield vs.\@ junction size. (c) Heat map of the average normalized resistance, $\overline{R_0}$. (d) Histograms of $R_0$ for four different junction areas. (e) $CV_{R_N}$ (points) vs.\@ junction area, $A$, and a decaying fit (continuous line) to equation (\ref{cov}). (f) $CV_A$ vs.\@ junction area, according to the equation $CV_A = 2\sigma_d/\sqrt{A}$.}}
    \label{fig:reproducibility}
\end{figure*}

\section*{Reproducibility of junction resistance}

The transmon qubit is essentially an anharmonic oscillator with fundamental transition frequency \cite{koch_charge-insensitive_2007}
\begin{equation}\label{freq_v2}
\centering
f_{01} \approx 1/(2\pi\sqrt{L_J C}) - E_C/h.
\end{equation}
Here, the charging energy, $E_C = e^2/(2C)$, depends on the qubit's total capacitance $C$, where $e$ is the electron charge.
The Josephson inductance, $L_J = \Phi_0/(2\pi I_c$), depends on the junction critical current, $I_c$. ($\Phi_0=h/(2e)$ is the magnetic flux quantum and $h$ is Planck's constant.)
In turn, $I_c$ is related to the junction normal-state resistance, $R_N$, via the Ambegaokar--Baratoff relation, \cite{ambegaokar_tunneling_1963} $I_c R_N=\pi\Delta/(2e)$. 
These equations indicate that $C$ and $R_N$ are the parameters that can influence the reproducibility of the qubit frequency across a wafer. The material and thickness-dependent gap parameter, $\Delta$, is not expected to fluctuate across a wafer at zero temperature. \cite{BCS, scaling_delta_2,scaling_delta}
The capacitance $C$ is dominated by a large planar capacitor with small fabrication-induced variation. Simulation shows that even 0.3~$\mu$m variation in the line-width of the capacitor changes the capacitance by $\sim$1$\%$. As a result, $R_N$ is the dominant parameter that causes variation in the qubit frequency, and following the analysis of ref. \onlinecite{Krantz2010InvestigationOT}, the deviation in $f_{01}$ is half that of $R_N$.

Statistical studies of normal resistance have been reported for both niobium- and aluminum-based JJs fabricated using different methods. Bumble et al. \cite{Bumble_rep} found a 2--4$\%$ on-chip resistance variation for circular Nb junctions (with AlO$_x$ tunnel barrier) of 0.33~$\mu$m$^2$ area. Tolpygo et al.
\cite{MIT_reproducibility_2015} found a value as small as 0.8$\%$ for larger-area junctions (1.8~$\mu$m$^2$), and 8$\%$ for 0.03-$\mu$m$^2$ junctions.  
Lotkhov et al. \cite{lokhtov_rep} reported 10--20$\%$ on-wafer spread for Al JJs with junction areas between 0.125 and 0.25~$\mu$m$^2$. 
Pop et al. \cite{pop_fabrication_2012} showed 3.5$\%$ variation on a single chip for Al JJs with junction areas ranging between 0.02 and 0.2~$\mu$m$^2$. 
More recently, a larger-scale reproducibility study over several wafers by Kreikebaum et al. \cite{kreikebaum2019improving}  showed an average on-chip variation of 1.8$\%$ and a wafer-scale spread of less than 3.5$\%$, although during subsequent fabrication of qubits, it increased to 6.9$\%$.

Using the PICT process, we fabricated thousands of test junctions and measured their resistance for a wafer-scale study of reproducibility. The 76-mm wafer included forty chips of size $0.5\times 0.5~\mathrm{cm}^2$, and each chip had 100 test junctions with 10 different sizes. The focus of this study was on small JJs (0.01 to 0.16~$\mu$m$^2$), the typical sizes used for transmon qubits. We measured the junction resistances using an automated probe station at room temperature (only measurements with coefficient of determination higher than 0.99 were considered). Figure~\ref{fig:reproducibility}(a) shows a heat map of the fabrication yield of each chip. The total wafer-scale yield is about 99.1$\%$. The bottom panel of Fig.~\ref{fig:reproducibility}(b) shows the yield as a function of junction size over the whole wafer.
 
The top panel of Fig.~\ref{fig:reproducibility}(b) shows the mean resistance for each junction size across the wafer, $\overline{R_N}(A)$, on a log scale with error bars representing one standard deviation. The continuous line is a linear fit with a slope of $\sim$-0.9, close to the expected number of $-1$, since $R_N \propto 1/A$. The deviation from the $-1$ slope is caused by the constant line-width bias (here $\sim\!28$~nm) in the EBL pattern compared to the CAD design. A heat map for the average normalized resistance, $\overline{R_0}$, of each chip is shown in Fig.~\ref{fig:reproducibility}(c). To obtain $R_0$ for one junction, its resistance is divided by the mean resistance of junctions with the same size across the wafer, such that $R_0 = R_N/\overline{R_N}$. The observed gradient of $\overline{R_0}$ over the wafer may be caused by uneven development and descumming (oxygen plasma). The inter-chip variation of the resistance is 3.7$\%$ across the wafer, with a best on-chip number of 1.4$\%$ and an average of 2.7$\%$. The variation of $R_N$ has a strong size dependence, \cite{MIT_reproducibility_2015} especially for small junctions, as shown in Fig.~\ref{fig:reproducibility}(d). The figure compares histograms of $R_0$ for four different junction areas across the wafer. Figure~\ref{fig:reproducibility}(e) shows, in red squares, the coefficient of variation (CV), the standard deviation of $R_N$ divided by the mean, as a function of the junction area for all junction sizes. 

Variation in $R_N$ can be caused by both the tunnel barrier thickness and the junction area $A$, since
\begin{equation}\label{resistance}
R_N = \frac{R_J}{A}, 
\end{equation}
where $R_J$ is the resistance per unit area of the junction. 
Assuming $R_J$ and $A$ are two independent variables, the CV of $R_N$ can be expressed as \cite{variance, Cov_appx}
 \begin{equation}\label{cov}
\centering
 CV_{R_N}^2 = CV_{R_J}^2 CV_{A}^2 + CV_{R_J}^2 + CV_{A}^2.
\end{equation}
In this equation, $CV_{R_J}$ is solely determined by the uniformity of the oxide barrier. On the other hand, to extract the dependence of $CV_{R_N}$ on $A$, we can derive an expression for $CV_A$ in terms of $A$ itself. Given the simple case of a square junction with side length $d$, $A = d^2$ and $\sigma_A = 2 d \sigma_d $, where $\sigma$ denotes the standard deviation. Dividing the latter equation by $A$, we obtain $CV_A = \sigma_A / A = 2 \sigma_d / \sqrt{A}$. Now, $\sigma_d$ is mainly determined by the lithography process, including exposure, development and descumming, and it is assumed to be a certain constant that does not scale with $d$. One can then fit the data in Fig.~\ref{fig:reproducibility}(e) to equation (\ref{cov}) after substituting for $CV_{A}$, and extract the constants $CV_{R_J}$ and $\sigma_d$ from the fit. 
We find $CV_{R_J} = 2.3\%$ and  $\sigma_d =  3$~nm. $CV_A$ as a function of the junction area is plotted in Fig.~\ref{fig:reproducibility}(f). 
This determination of $CV_A$ and $\sigma_d$ was done using the nominal, designed junction area; however, we can improve the accuracy by taking into account the previously determined 28~nm line-width bias. In this way, we find $CV_{R_J} = 1.8\%$ and  $\sigma_d =  4$~nm. For the typical JJ sizes (0.02---0.06~$\mu$m$^2$) used for fixed-frequency transmon qubits, $\sigma_d=4$~nm corresponds to $CV_{A}$ of 2.9---4.6$\%$. In this case, 60---70$\%$ of the total variation in $R_N$ is attributed to fluctuations in the junction area, while the rest is attributed to the inhomogeneity of the tunnel barrier.

Improving the reproducibility of $R_N$ requires minimizing the two parameters $CV_{R_J}$ and $\sigma_d$. For $CV_{R_J}$, it was shown that the uniformity of the AlO$_x$ barrier heavily relies on the uniformity and the morphology of the underlying Al layer, in addition to the oxidation conditions. \cite{Fritz_KIT} For $\sigma_d$, the lithographic process is the main contributor. A high-resolution resist and an optimized EBL process, in addition to an improved recipe of resist development and descumming, can lead to a minimal deviation in the feature size. \cite{kreikebaum2019improving,2inch}

To summarize, we proposed and demonstrated a simplified process to fabricate a Josephson junction and its patch layer that provides a superconducting, galvanic connection of the junction to the circuit. The process relies on shadow evaporation from three angles and fabricates the junction and the patch in only one lithography step. Suitable for making superconducting qubits, our method reduces the total number of junction fabrication steps by half without introducing further losses. Moreover, we statistically studied the reproducibility of the junctions and achieved a high fabrication yield. The junctions' resistance variation showed a strong dependence on the width, with an average variation of less than 3.7$\%$, comparable to the best values that are reported by other research groups. The variation can be reduced by optimizing the lithography process, and by improving the uniformity of the tunnel barrier.

\section*{Acknowledgements}
This research was funded by the KAW Foundation through the Wallenberg Center for Quantum Technology (WACQT) and by the EU Flagship on Quantum Technology H2020-FETFLAG-2018-03 project 820363 OpenSuperQ. The authors acknowledge the use of Nanofabrication Laboratory (NFL) at Chalmers and thank the staff, especially Henrik Frederiksen, Mats Hagberg, Bengt Nilsson and Johan Andersson. We also acknowledge Lars Jönsson for machining the sample boxes, and Giovanna Tancredi for her valuable feedback on the manuscript.

\section*{References}
\bibliography{references_APL}

\begin{thebibliography}{37}
\bibitem [{2in()}]{2inch}%
  \BibitemOpen
  \href@noop {} {}\bibinfo {note} {We would like to note that we performed the
  same reproducibility study on a 50-mm wafer and achieved the same yield.
  However, the wafer-level variation was higher since the EBL system was not
  able to accurately focus for patterns close to the rim of the wafer. On a
  50-mm wafer, the average wafer-scale variation was 7\%. However, an inner
  area on the wafer, containing 25 chips, showed a variation of 3.5\%
  only.}\BibitemShut {Stop}%
\end{thebibliography}

\end{document}